# Effects of Intermediate Frequency Bandwidth on Stepped Frequency Ground Penetrating Radar

Wenhao Luo, Hai-Han Sun, Yee Hui Lee,
Abdülkadir C. Yücel
School of Electrical & Electronic Engineering
Nanyang Technological University, Singapore
wenhao.luo@ntu.edu.sg, haihan.sun@ntu.edu.sg,
eyhlee@ntu.edu.sg, acyucel@ntu.edu.sg

Genevieve Ow and Mohamed Lokman Mohd Yusof
Centre for Urban Greenery & Ecology
National Parks Board, Singapore
GENEVIEVE_OW@nparks.gov.sg,
Mohamed_Lokman_Mohd_Yusof@nparks.gov.sg

*Abstract*—A stepped frequency ground penetrating radar (GPR) system is used for detecting objects buried under high permittivity soil. Different intermediate frequency bandwidth （IFBW） of the mixing receiver is used and measurement results are compared. It is shown that the IFBW can affect the system's signal-to-noise ratio (SNR). Experimental results show that objects of different materials can clearly be detected when the appropriate IFBW is used.

I. INTRODUCTION

The humidity in a tropical country is generally high, especially during the rainy season. This high humidity in the environment may result in high soil moisture content. The high permittivity of high moisture soil generally degrades the performance of the GPR [1]. Therefore, to improve the detection capability of the GPR in such an environment, suitable system parameters should be investigated.

There has been several research performed on time-domain pulse GPR. In [2], a time-domain pulse GPR is reported to have low detection resolution due to the hardware capability that limits its pulse width. Compared to the time-domain pulsed GPR, the stepped frequency GPR is more suitable for this research for several reasons. Firstly, it is easy to set the desired frequency components with equal power levels. Secondly, the frequency bandwidth of the system adapts the signal bandwidth to match the antenna well; finally, it is convenient for signal processing since the transmitted and received signals are both in the frequency domain [3].

In this research, two objects with different materials are buried under the high moisture soils. These objects are a tree trunk that has been soaked in water for more than a week and an empty aluminum water container. Intermediate frequency bandwidth (IFBW) can be used to control the signal-to-noise ratio of the received signal, the smaller the IFBW, the better the signal-to-noise ratio (SNR) [4]. In the experiment, we study the effect of different IFBW values on the detection capability of underground objects. The most suitable IFBW value is obtained after comparing the experiment results.

II. MEASUREMENT OF DIELECTRIC PARAMETERS

The soil in the testbed is prepared based on the soil composition ratio of urban parks provided by National Parks Board, Singapore. It is comprised of a mixture of 3: 1: 1 ratio of topsoil, sand, and organic matter. The testbed is then placed in the natural environment.

Since the GPR system has a wide operating bandwidth, it is important to know the permittivity and conductivity of the soil over the broad frequency range. The measurements of the soil properties were carried out using the Agilent 85070D Dielectric Probe [5], which measures the real and imaginary parts of the relative permittivity. The measured results are shown in Fig. 1. The average relative permittivity is $\varepsilon_c = 27.5 + j2.5$.

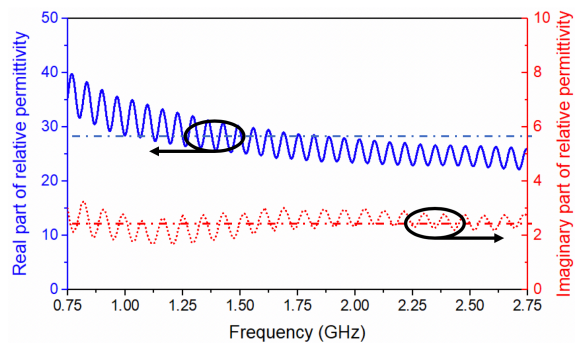

Fig. 1 Measured real and imaginary parts of the relative permittivity of soil.

When measuring the soil in the testbed, the entire testbed was divided into 12 equal parts, and 3 points were tested for each part. We found that the permittivity value of the soil near the edge of the testbed is significantly different from that in the middle, which can be attributed to the influence of the sidewalls. The results shown in Fig. 1 are the averaged value of 18 points that are collected in the middle of the testbed.

One of the tested objects is a wet V-shape tree branch, as shown in Fig. 2(a). In order to imitate a tree trunk, the dry tree brunch was soaked in water for more than one week to increase its water content. By doing so, its permittivity will be similar to the value of the soil and the detection capability should be reduced. The other test object is an aluminum bottle with the same size as the trunk, as shown in Fig. 2(b). This provides a reference as an object that is relatively easy to be detected.

This work is funded by National Parks Board, Singapore.

The characteristics of the wet trunk and aluminum bottle are also tested using the Agilent 85070D Dielectric Probe, and the averaged results of the test results are shown in Table I.

### III. LABORATORY EXPERIMENTS

The two buried objects are detected using a bi-static GPR system with two different IFBW values: 50 Hz and 500 Hz. The measured B-scan results are plotted for analysis and comparison.

### A. GPR system setup

The GPR system is made of two identical (EMCO/ETS-Lindgren 3115) double-ridged waveguide horn antennas operating from 0.75 GHz to 18 GHz. The antennas are placed parallel to each other with a center-to-center distance of 30 cm and are placed above the soil surface with a distance of 5 cm. They are connected to ports 1 and 2 of a Rohde & Schwarz ZVL6 Vector Network Analyzer (VNA) respectively as the transmitter and receiver. The key set-up parameters of the VNA are listed in Table II. The objects are buried at a depth of about 8cm to 10 cm. The S-parameters ($S_{21}$) are acquired at a space interval of 2 cm along the testbed. Each test point generates an A-scan. B-scan images can be obtained by combining several A-scans along a measurement line.

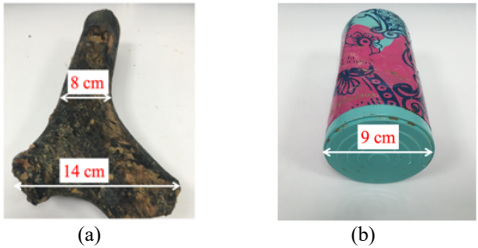

(a)　　　　　　　　(b)

Fig. 2 Pictures of (a) Water-soaked tree brunch, and (b) Aluminum bottle.

TABLE I. MEASURED CHARACTERISTICS OF BURIED OBJECTS

| Object | $\varepsilon_{real}$ | $\varepsilon_{imaginary}$ |
|---|---|---|
| Wet trunk | 26.01 | 5.73 |
| Aluminum bottle | 7.12 | 1.13 |

TABLE II. SUMMARY OF SETUP OF ROHDE & SCHWARZ ZVL6 VNA

| Parameters | Value |
|---|---|
| Number of steps | 4001 |
| Start frequency (GHz) | 0.75 |
| Stop frequency (GHz) | 2.75 |
| Tx Power (dBm) | -10 |
| I.F. bandwidth (Hz) | 50 |
|  | 500 |

### B. Experimental results

Fig. 3 shows the measured B-scan results after background removal operation. Since the placement of the antenna is completely symmetrical with respect to the testbed, after the background removal operation, the effect of the side walls of the testbed on the experimental results is minimized. The highlighted points that indicate the objects are clear, from which we can estimate the location of the objects. From Fig. 3(a), a V-shape can be observed, which is similar to the shape of the tree brunch. Also, it can be found that the highlighted area representing the branch is more concentrated for an IFBW of 50 Hz. This is because a smaller IFBW leads to a higher SNR. In Fig. 3(b), the advantage of low noise interference with IFBW of 50 Hz is more obvious. In this case, the hyperbola shape of the B-Scan can easily be distinguished from the surrounding noise.

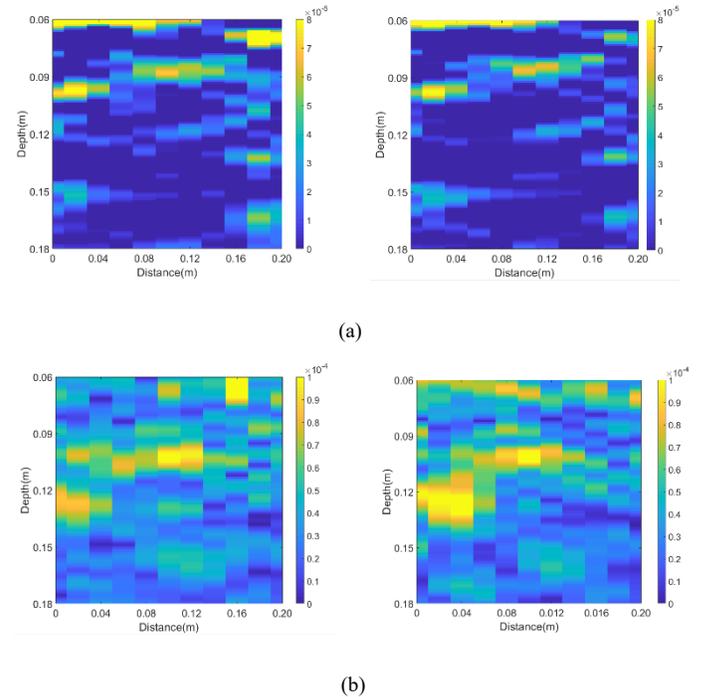

Fig. 3 Experimental results: IFBW = 50 Hz (left) and IFBW = 500 Hz (right). (a) water-soaked trunk, and (b) Aluminum bottle.

### IV. CONCLUSION AND PERSPECTIVES

The paper presents an evaluation of the impact of IFBW on the detection capability of buried objects using GPR. In our study, a bi-static GPR system has been successfully operated to collect B-scans of two different objects. From the results, the stepped frequency GPR with lower IFBW yields better detection capability with less noise and higher resolution. Further investigation into the IFBW and other system parameters for better resolution will be performed.